\documentclass[twocolumn,aps,amsmath,amssymb,showpacs]{revtex4}
\pdfoutput=1
\usepackage{graphicx}
\usepackage{hyperref}

\def\de{\delta}

\def\eps{\varepsilon}

\def\La{\Lambda}

\def\beq{\begin{equation}}
\def\eeq{\end{equation}}
\def\bea{\begin{eqnarray}}
\def\eea{\end{eqnarray}}

\begin{document}
\title{Apparent evidence for Hawking points in the CMB Sky}
\author{Daniel An$^{1}$, Krzysztof A. Meissner$^2$, Pawe{\l} Nurowski$^3$ and Roger Penrose$^4$}
\affiliation{$^1$Science Department SUNY Maritime College,\\
 6 Pennyfield Av., New York 10465, USA\\
$^2$Faculty of Physics, University of Warsaw\\
Pasteura 5, 02-093 Warsaw, Poland\\
$^3$Center for Theoretical Physics of PAS\\ 
Al. Lotnik{\'o}w 32/46, 02-688 Warsaw, Poland\\
$^4$Mathematical Institute, Oxford University,\\
Radcliffe Observatory Quarter,\\
Woodstock Rd., Oxford OX2 6GG, UK
}

\vspace{3mm}

\begin{abstract}
\centerline{\bf Dedicated to the memory of Stephen Hawking}
\vspace{3mm}
\noindent
This paper presents strong observational evidence of numerous previously unobserved anomalous circular spots, of significantly raised temperature, in the CMB sky. The spots have angular radii between 0.03 and 0.04 radians (i.e. angular diameters between about 3 and 4 degrees). There is a clear cut-off at that size, indicating that each anomalous spot would have originated from a highly energetic point-like source, located at the end of inflation -- or else point-like at the conformally expanded Big Bang, if it is considered that there was no inflationary phase. The significant presence of these anomalous spots, was initially noticed in the Planck 70 GHz satellite data by comparison with 1000 standard  simulations, and then confirmed by extending the comparison to 10000 simulations. Such anomalous points were then found at precisely the same locations in the WMAP data, their significance confirmed by comparison with 1000 WMAP simulations. Planck and WMAP have very different noise properties and it seems exceedingly unlikely that the observed presence of anomalous points in the same directions on both maps may come entirely from the noise. Subsequently, further confirmation was found in the Planck data by comparison with 1000 FFP8.1 MC simulations (with $l \leqslant 1500$). The existence of such anomalous regions, resulting from point-like sources at the conformally stretched-out big bang, is a predicted consequence of conformal cyclic cosmology (CCC), these sources being the Hawking points of the theory, resulting from the Hawking radiation from supermassive black holes in a cosmic aeon prior to our own.

\vspace{0.3cm}
\noindent
{\bf Key words:} cosmic background radiation -- cosmology.
\end{abstract}

\pacs{04.20.Ha,04.70.Dy,98.80.Bp,98.80.Ft}
\maketitle

\vspace{3mm}
{\bf 1. Conformal Cyclic Cosmology}

\noindent
The proposal of conformal cyclic cosmology (CCC) (Penrose 2006, Penrose 2010, Penrose 2018)
provides an alternative to the initial inflationary epoch of current cosmology, 
where CCC replaces inflation by the $\La$-driven exponential expansion  
of a pre-Big Bang universe epoch, referred to as a previous {\it aeon}; compare (Gasperini $\&$ Veneziano 1993). $\La$ is taken as an absolute positive (cosmological) constant 
and the entire universe history is taken to be an unending sequence of such aeons, each 
beginning with a big bang and ending with a $\La$-driven exponential expansion. 
The conformal infinity (see (Penrose 1964)) of each aeon joins conformally smoothly to the conformally expanded big bang origin of the subsequent aeon.

The conformal infinity of each aeon is spacelike because $\La >0$ (Penrose 1965) and the conformally stretched big bang of each is also spacelike, so the matching has some geometrical rationale. Moreover this identification is physically as well as geometrically plausible. The conformal squashing of the exceedingly cold and rarefied remote future results in an enormous increase in the temperature and density. Correspondingly, the conformal stretching of each big bang provides an enormous reduction in the temperature and density, so the conformal matching is not physically implausible. Moreover, the matter content of the remote future of each aeon is essentially dominated by photons, these being governed by the conformally invariant Maxwell equations so that, with regard to this dominant matter component of the remote future, this conformal picture of space-time is physically appropriate. Correspondingly, as we proceed back in time into the big bang of each aeon, we find that massive particles attain kinetic energies so large that their masses become physically irrelevant and act as conformally invariant massless entities in the limit as that aeon's big bang is approached. It is thus argued that the conformal joining of each aeon's conformal future infinity to the conformally stretched big bang of a subsequent aeon may be considered as making both geometrical and physical sense.

There are two additional issues concerning the remote future of each aeon. One of these is the presence of a certain proportion of massive particles, atoms, or molecules, that one could expect to survive to the indefinite future of each aeon, such as hydrogen atoms, protons (and/or positrons), electrons, and neutrinos (or neutrons, etc. in neutron stars), which in current theory retain their mass indefinitely. This is dealt with in CCC by the proposal of a $\La$-driven ultimate mass fade-out, where it is argued that the masses of all particles asymptotically fade to zero in the remote future of each aeon. One possible rationale for this is, with $\La$ being fundamental, that the most basic group of particle physics should be the de Sitter rather than Poincar{\'e} group, and that mass, not being a Casimir operator of the de Sitter group, need not be an absolute constant, for a stable particle, in the presence of $\La$ (see (Penrose 2018)). Charge remains constant in CCC, as does $\hslash$, from which it follows that the ground state of hydrogen will gradually dissociate as the electron mass and proton mass gradually fade away. Protons themselves might decay, but since charge is conserved one may expect that the ultimate positively charged decay product would be positrons, subject, as with electrons, to ultimate asymptotic mass fade-out.

This proposal of ultimate mass fade-out is indeed an assumption of CCC.  The remaining remote-future issue is the ultimate fate of black holes. Accepted theory (Hawking 1974, Hawking 1975) asserts that these will eventually evaporate away by Hawking radiation as the universe finally cools to near absolute zero. CCC is in accordance with this view, whose remarkable consequences will be discussed further in Section 3. For now, it may be remarked, first, that galactic clusters remain essentially bound as the universe (exponentially) expands, and it may be expected that the majority of the mass in a cluster should eventually be swallowed by a hugely supermassive black hole of perhaps up to some $10^{14}$ solar masses. The expectation is (Page 1976) that although it might take up to some $10^{106}$ years, virtually the entire mass of that black hole, and therefore of that galactic cluster, will ultimately be radiated away in the form of photons, or perhaps other particles that will have become effectively massless due to mass fade-out. Because of the extremely late stage, in each aeon, of this energy release, and owing to the huge conformal squashing involved, it all becomes effectively concentrated at a single point H (as marked in Fig.1.) of the crossover 3-surface X (as marked in Fig.1.) which joins that aeon's conformal infinity to the conformally stretched big bang of the subsequent aeon. We refer to such a point H as a {\it Hawking point}.

\begin{figure}[t]
\hspace{0cm}
\centering
\includegraphics[trim=0 5 0 0,clip,height=9cm]{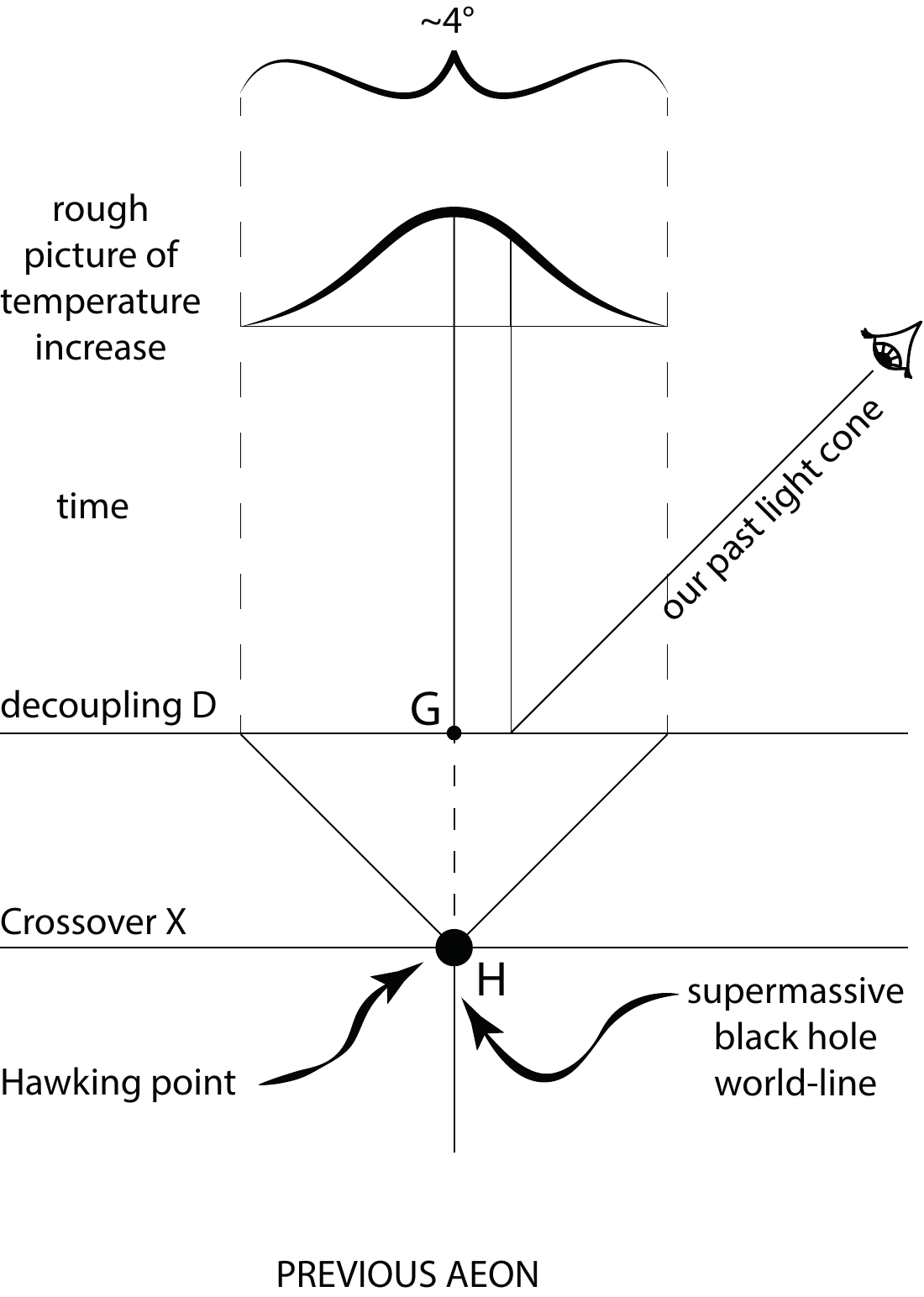}
\caption{The lower part is a conformal diagram representing the effect of a highly energetic event occurring at the space-time point H. In CCC, H is taken to be a Hawking point, where virtually the entire Hawking radiation of a previous-aeon supermassive black hole is concentrated at H by the conformal compression of the hole's radiating future. The horizontal line at the bottom stands for the crossover surface dividing the previous cosmic aeon from our own and describes our conformally stretched Big Bang. In conventional inflationary cosmology, X would represent the graceful exit turn-off of inflation. In each case, the future light cone of H represents the outer causal boundary of physical effects initiated at H, and such effects can reach D only within the roughly 0.08 radian spread indicated at the top of the diagram.
}
\label{fig}
\end{figure}

Before coming to the observational implications of this picture, it will be helpful to explain certain other aspects of CCC. It is important, first, to understand that CCC is not, in any major way, in conflict with conventional $\La$-CDM cosmology, from some $10^{-35}$ s (the presumed turn-off moment of inflation, in conventional theory) after our Big Bang (where the capitalized “Big Bang” refers to the specific moment that initiated our own particular aeon). From this moment onwards, within our current aeon, CCC is in basic agreement with the conventional $\La$-CDM picture -- though the Hawking points would modify things in a way that we argue here has some direct observational support. CCC's ultimate mass fade-out does not affect this agreement. The presence of Hawking points adds an intriguing new ingredient to the conventional picture, though not grossly altering it. Most particularly, CCC would not appear to alter, in any major way, the conventional description of the $\sim$380000-year period following the turn-off of inflation (graceful exit) and decoupling, being beautifully confirmed by agreement with the Planck-data power spectrum for $l$-values greater than about 40.

The actual input from inflation to this $\sim$380000-year evolution, in standard cosmology, is fairly minimal, but its role in providing a source for the almost scale-invariant temperature variations in the CMB needs an alternative explanation in CCC. This is provided by the $\La$-driven exponential expansion of the previous aeon, where inflation theory's inflaton quantum fluctuations are replaced by the effects, coming through from the previous aeon, of gravitation wave impulses that are the products of erebon decay in that previous aeon, erebons being the particles that constitute dark matter in CCC. Such dark matter is a necessary consequence of the equations of CCC (Penrose 2010, Gurzadyan $\&$ Penrose 2013, Penrose 2018). With a decay half-life of the order of $10^{11}$ years, both the near scale invariance of the CMB temperature fluctuations and the spectral index can find a CCC explanation. Details will be published elsewhere, but for a preliminary discussion, see (Penrose 2018).

It should also be pointed out that the arguments in support of inflation that it supplies an explanation for the temperature fluctuations that would appear to be acausal in a cosmology without inflation are easily explained within CCC . The existence of an aeon prior to ours gives ample scope for such correlations. For example, the gravitational wave signals considered in the following section would certainly provide correlations of this apparently acausal kind.\\

{\bf 2. Observable effects of previous-aeon black-hole encounters}

There are two quite distinct types of event concerning supermassive black holes in the previous aeon, that should be observable, according to CCC, and it is important not to confuse them. One of these would be the effects of gravitational waves coming from inspiralling pairs of supermassive black holes (discussed in this section) and the other the effects of the evaporation of supermassive black holes (discussed in the next section). The observable signals for the first type would be narrow-width circular rings, within which would be the observed effect of the final burst of this radiation, containing almost the entire energy emitted. These rings would often occur in concentric sets since several such events could often take place within the history of a single galactic cluster, whose extended world-line would reach the crossover surface at the central point of the rings. Such rings wold be distinguished by having anomalously low temperature variance around them (Gurzadyan $\&$ Penrose 2013, appendix B), or of significantly differing average temperature around the ring from that of a neighbouring ring concentric with it. Such rings could be of large angular diameter across the sky, but not of greater than around $40^\circ$ (Tod 2012, Nelson $\&$ Wilson-Ewing 2011).
Evidence for the existence of such rings in the WMAP data was found by Gurzadyan and Penrose (Gurzadyan $\&$ Penrose 2013) and by Meissner, Nurowski and Ruszczycki (Meissner, Nurowski $\&$ Ruszczycki 2013) in the WMAP data, the latter providing a 99.7\% statistical confidence level for the reality of the effect. Subsequently, An, Meissner and Nurowski  (An, Meissner $\&$ Nurowski 2018) examined the Planck satellite data, again finding a significant effect, at a 99.4\% confidence level.

The WMAP analysis provided in (Gurzadyan $\&$ Penrose 2013) was repeated by DeAbreu, Contreras, and Scott (DeAbreu, Contreras, $\&$ Scott 2015), and although a virtually identical picture of the circular features was obtained, these authors argue that the presence of such features is not statistically significant. They did not, however, repeat the further analysis actually given by Gurzadyan and Penrose, which showed that when the analysis was modified so as to search for somewhat elliptical rather than circular rigs, the numbers drop dramatically, and continued to drop consistently, with increasing ellipticity, showing that concentric circular rings predicted by CCC have a preferential significance. The procedure adopted here is to use the same concentric-circle search as before, but applied to a CMB sky that is twisted about the galactic polar axis, this being equivalent to a search for concentric elliptical shapes in the real sky. This procedure is misunderstood by DeAbreu, et al, who argue that the temperature variance profile of the twisted sky is different from that of the real sky, a point which is actually irrelevant to the analysis of (Gurzadyan $\&$ Penrose 2013), which depends only on that for the real sky. We should also point out that other earlier claims (such as (Wehus $\&$ Eriksen 2011, Hajian 2011)) that they do not find statistically significant evidence for the signals, later argued for more forcefully by Gurzadyan and Penrose (Gurzadyan $\&$ Penrose 2013), are not demonstrations that such signals do not actually exist, but indicate that a more sophisticated analysis is needed, such as that of (Gurzadyan $\&$ Penrose 2013, Meissner, Nurowski $\&$ Ruszczycki 2013, An, Meissner $\&$ Nurowski 2018).

More importantly, DeAbreu, et al. provide no explanation for the striking departure from isotropy in the observed features, where they comment only on the relatively minor feature that more centres are found in the galactic southern than in the northern hemisphere. However, the departure from isotropy is far more striking than this, which is made particularly obvious in the search for centres of triples of low-variance rings in the Planck satellite data as displayed in Fig. 2 of (Gurzadyan $\&$ Penrose 2016), which exhibits the centres of concentric triples (at least) of low-variance circular rings. We notice that the vast majority (in fact around 83\%) of the 1134 centres found lie within three roughly elliptical regions (two in the southern and one in the northern galactic hemisphere), each of not more than about 3\%
of the entire region of the sky being examined.

It should be pointed out that such anisotropy/inhomogeneity is not a prediction of CCC, but is perfectly consistent with that theory. The trouble lies more with the conventional inflationary picture, according to which, such gross departures from uniformity are hard to explain. In the CCC picture, there is no inflation, and therefore no strong requirement that the large-scale spatial universe be particularly homogeneous or isotropic, and the strong clumping depicted in Fig. 2 of  (Gurzadyan $\&$ Penrose 2016) could be explained by the presence, in the previous aeon, of enormously large superclusters of galaxies in the directions of these three concentrated regions. Such an interpretation is enhanced by the fact that there is also a strong “clumping” in the colour coding in Fig. 2 of (Gurzadyan $\&$ Penrose 2016), a feature whose strong significance for CCC, and also its relevance to Hawking points, will be discussed elsewhere.\\

{\bf 3. Hawking points in CCC}

We now come to the essential purpose of this paper: the observational implications of the ultimate fate of supermassive black holes in the previous aeon, and the observational implications for our current aeon. The picture that we should bear in mind is that whereas superclusters gradually disperse, owing to the  exponential expansion of the later stages of each aeon, individual galactic clusters remain bound. The supermassive black holes within the galaxies of each cluster, will begin to feel each other out and then spiral into one another to form a single hugely massive black hole. As time evolves, that remaining supermassive black hole would be expected to swallow perhaps the majority of the matter in that galactic cluster, though some fair portion might escape into outer space, to join intergalactic hydrogen and dark matter already there. Eventually, the mass content of the aeon would consist, to large extent, of huge supermassive black holes of perhaps up to $10^{14}$ solar masses, which would last for periods of up to perhaps $10^{106}$ years (Page 1976) after which it should have Hawking-evaporated away completely. 

We must bear in mind that, despite Hawking radiation being of an absurdly tiny temperature, over
the aeon's history the entire mass-energy of the hole will be finally radiated away and, in the conformal picture,
this will take place within what would effectively be a single Hawking point H,
only infinitesimally beneath X. Being mainly in the form of photons (and some neutrinos and other mass-faded particles) this
radiation comes directly through X to heat the initial material in the succeeding aeon enormously, just to the future
of the Hawking point H, depositing the hole's entire mass-energy there. This hugely heated region would then gradually spread out in our aeon until reaching the decoupling 3-surface D, providing something like a Gaussian distribution on D, centred at a point G on D, just to the future of H in the conformal picture, where the spread, duly constrained
by the speed of light, should, from our vantage point, be no greater than the maximum causally allowed
diameter of around 4$^\circ$ (i.e. a radius $\le 0.035$ radians). What we see from our current vantage point would be the intersection
of our past light cone C with this small distribution on D, which appears to us as a small Gaussian-like distribution centred
at the Hawking point H, having spread out to subtend an angular distance of no greater than $0.035$ radians on either side of G 
(i.e. within H's light cone), see Fig.1. 

The actual temperature profile depends on the detailed particle physics involved, in accordance with conventional theory, and, for the enormously highly energetic particles involved, we would expect that it should appear to us like a Gaussian with maximum temperature at at G, and cooling off as it spreads out to something a bit less than the causally allowed maximum of 2$^\circ$ radius -- less, because C would normally not quite pass through G (see Fig 1.)  We call such a (necessarily circular) region of raised temperature a {\it Hawking disc}, the presence of such features in the actual CMB sky being a critical issue for CCC. In Section 4, we describe the details of our search for Hawking discs in the actual CMB where, according to the above discussion, we take note of the fact that the temperature of a Hawking disc ought to have a maximum at its centre, and which falls off in a Gaussian-like way towards its outer edge. In order to identify such discs we look for annular regions in the CMB sky, taken to be concentric with such a proposed Hawking  disc, and see if we find a significant drop in temperature from the inner to outer boundary of the annulus. The test is to see whether these features are found in the actual CMB sky with parameter values as are predicted by the theory described above.

Accordingly, we should indeed expect to find such annuli, but where the outer edge has a radius no larger than about 0.035 radians, and where we anticipate a sharp cut-off in numbers for such discs having a greater outer radius than this, and not so many of them with outer radii of less than around 0.02 radians. These are expectations that we infer from Fig.1. As we shall see in Section 4, our findings appear to be strongly in accordance with these expectations.\\

{\bf 4. Anomalous small regions in the CMB map}

\noindent

In order not to prejudice our search in favour of the particular signals anticipated by CCC as described above, we broadened our search, so as to be for annuli with inner radii $r_1$ of between zero and 0.04 radians and outer radii $r_2$ from 0.01 up to 0.08 radians. The annular width  $\eps= r_2 - r_1$ is taken to run from 0.01 to 0.04 radians, all taken in steps of 0.01 radians. Moreover we allow for the temperature slope from inner to outer radius to be positive, rather than the CCC-expected negative slopes.

Remarkably, we find, at a 99.98\% confidence level (of agreement between this particular aspect of CCC theory and the Planck satellite data) the existence of this previously unnoticed family of anomalously energetic small circular regions in the CMB sky. A comparison with 1000 conventional Planck simulations provided a powerful case for the possibility of there being actual signals of the kind that would be consistent with the theoretical considerations that we have described earlier in this paper, the signal appearing to be outstandingly strong. Accordingly, we compared our CMB findings with a further independent 9000 simulations. Again we found the same strong signal as before, at the same parameter values, and our confidence level of better than 99.98\%, referred to above, is based solely on these subsequent 9000 simulations.

To provide additional confirmation of this result we also used 1000 FFP8.1 MC simulations (with $l\leqslant 1500$) and obtained the same result. As an independent test, we compared the WMAP `deconvolved' real map with 1000 simulations and again arrived at the same result, noting that the five strongest anomalous regions of the Planck map were found to occupy precisely the same locations as five of the strongest locations found in the WMAP data.  Planck and WMAP have very different noise properties and it seems unlikely that the observed presence of anomalous points in the same directions in both maps could come entirely from the noise.

The details and the particular motivations underlying our current search are given below. In this section we simply provide a brief description of the particular anomalous regions of the CMB sky that we appear to be seeing, and raise some of the significant implications of this. What our search reveals is a multitude of distinctive circular spots in the CMB, of increasing temperature towards their centres, having an intensity somewhat more than an order of magnitude greater than the standard $10^{-5}$ temperature fluctuations.

A striking and noteworthy feature of these anomalous spots is that within the range of diameters that we examine, there is a sharp cut-off at an angular diameter of around $0.08$ radians. To understand the puzzle for conventional inflationary theory that is raised by this finding, it is helpful to examine the conformal diagram of Fig. 1. This depicts the 380000-year period between a powerful source of energy at the space-time point H and the decoupling surface D. We note that if the source had not been in the extremely early universe, as depicted in the figure, but within the following 380000 years before reaching D, then the spreading out of the signal at D could indeed be constrained to within around $0.08$ radians in the observed CMB. But the physics within that 380000 year period is well understood, being superbly confirmed by the close agreement with the CMB power spectrum at $l$ values larger than about 40, so it is hard to see how the signals we see could originate in this way. On the other hand, any hugely energetic disturbance that took place much earlier than the turn-off of inflation (the so-called graceful exit moment) would have spread to a far larger diameter when reaching D. In conventional inflationary theory, unless H were indeed constrained to be very close to the turn-off surface of inflation represented by the horizontal line X, the spread of this energy to the future of H would be expected to be much larger than the observed angular diameter of around $0.08$ radians. Accordingly, H would have to be very close to the turn-off of inflation, as depicted in the figure, which would seem to be problematic for inflationary theory. On the other hand, as has been argued above, the expectations of CCC are well in accordance with these observations and, indeed, such features are predictions of that theory.\\

{\bf 5. Details of the analysis} 

\noindent
In the present paper the procedure we use to analyse Planck and WMAP data is similar to the one used by us before in (Meissner, Nurowski $\&$ Ruszczycki 2013) and (An, Meissner $\&$ Nurowski 2018) (Galactic equatorial belt excluded, imposed masks etc.) with one crucial difference. To explain this difference we recall that in the previous searches (i.e. looking for the ring-type structures) the assumed profile consisted of two contiguous concentric annuli, the inner with negative weight and the outer with positive weight. The convolution of the profile (with different angular radii and width of the annuli) with the actual temperature $T$ was calculated for rings in different directions in the sky. Such calculations were performed both for the real maps as measured by WMAP and Planck (70 GHz, SMICA, SEVEM...) and, initially, for 1000 artificial maps generated with the observed CMB power spectrum. Then the CDF’s of the results were compared using the procedure described in (Meissner 2012). The results showed that in some cases {\it none} of the artificial maps out of this 1000 performed better than the real map. 

The difference between our previous approach and the present one is that now we are looking for the slopes of T around a given direction for an annulus of inner angular radius $r_1$ and width $\eps$. The slope in a given direction is calculated by the formula (minimizing $\sum(\de T_i-a\,x_i-b)^2$ with respect to $a$ and $b$)
\beq
a=\frac{n\sum (x_i\, \de T_i)-(\sum x_i)(\sum \de T_i)}{n\sum x_i^2-(\sum x_i)^2}
\eeq
where $x_i$ is the angular distance of point $i$ from the given direction, $\de T_i$ is the temperature at this point and $n$ is
the number or points in the annulus. The sums run over all points in an annulus around the given direction. Having the slopes for
all directions on the sky we create the CDF for a given map. Then we create a “theoretical” CDF by averaging
the CDFs of all these 10000 artificial maps and use the formula from (Meissner 2012) to calculate 2 numbers describing extremal,
positive and negative, parts of CDF. Then we check how many artificial maps “outperform” the real map,
separately for positive and negative slopes. 

Table I, below, gives the results for smallest annuli. Values $r_1$ for the inner radii are given in column 1, and for widths $\eps$, in columns 2 and 7. The 3$^{\rm rd}$ column shows the number $N_-^{1k}$ of artificial maps  (out of this 1000) outperforming the real map with large positive slopes and the 4$^{\rm th}$ and 9$^{\rm th}$ columns, the number  of artificial maps (out of this 1000) outperforming the real map with large negative slopes. The zeroes in columns 4 and 9 show that for annuli of widths $\eps=0.02$ or $\eps=0.03$ (with inner radius $r_1=0.01$) there are no artificial maps outperforming the real map with large negative slopes i.e. with the temperature decreasing outwards. The appearance of these zeroes led us to perform an analysis with a much larger set of maps (9 000 more, in addition to the original 1000) either to confirm or reject the hypothesis of the physical validity of the strong signal seen in the true data, for $r_1=0.01$ and widths either $\eps=0.02$ or $\eps=0.03$ both absent in the first set of 1000 artificial maps. We see in the 11$^{\rm th}$ and	
6$^{\rm th}$ ($N_-^{10k}$) columns that the hypothesis is confirmed and we can estimate the probability of the purely random appearance of annuli ($0. 01,\  0. 02$) as approximately $0.01$\% and ($0. 01,\ 0. 03$) as approximately $0.02$\%, providing us with a confidence level of better than $99.98$\% that there is a genuine signal in the true CMB at  $r_1=0.01$ and widths at either $\eps=0.02$ or $\eps=0.03$.

\begin{center}
\begin{table}[h]
\caption{\# of artificial maps outperforming the real Planck 70 GHz map}
\vspace{3mm}
\begin{minipage}[b]{0.45\linewidth}\centering
\begin{tabular}{|c||c|c|c|c|c|}
\hline
  $r_1$& $\eps$ &  $N_+^{1k}$ & $N_-^{1k}$&$N_+^{10k}$&$N_-^{10k}$\\\hline\hline
 0.0&0.01& 921 & 242&9358&2406\\\hline
0.01&0.01& 952 & 139&9592&1422\\\hline
0.02&0.01& 215& 831&2199&8330\\\hline
0.03&0.01& 625 & 905&6398&9062 \\\hline
0.04&0.01& 182& 910 &1926&9310\\\hline
\end{tabular}
\end{minipage}
\hspace{5mm}
\begin{minipage}[b]{0.45\linewidth}\centering
\begin{tabular}{|c|c|c|c|c|}
\hline
  $\eps$ &  $N_+^{1k}$ & $N_-^{1k}$&$N_+^{10k}$&$N_-^{10k}$\\\hline\hline
0.02&710 & 186 &7110&1700\\\hline
0.02& 734 & 0&7232& 1 \\\hline
0.02& 384 & 110 &4056&1119\\\hline
0.02& 258 &978 &2627&9810\\\hline
0.02& 991& 921 &9929&9169\\\hline
\end{tabular}
\end{minipage}
\end{table}
\end{center}

\vspace{-17mm}

\begin{center}
\begin{table}[h]
\begin{minipage}[b]{0.45\linewidth}\centering
\begin{tabular}{|c||c|c|c|c|c|}
\hline
  $r_1$& $\eps$ &  $N_+^{1k}$ & $N_-^{1k}$&$N_+^{10k}$&$N_-^{10k}$\\\hline\hline
0.0&0.03& 681 & 63 &6649&595\\\hline
0.01&0.03& 875 & 0 &8904& 2\\\hline
0.02&0.03& 756 & 601 &7567 &5855\\\hline
0.03&0.03& 180 & 666 &1979&6602\\\hline
0.04&0.03& 162 & 412 &1664 &4271\\\hline
\end{tabular}
\end{minipage}
\hspace{5mm}
\begin{minipage}[b]{0.45\linewidth}\centering
\begin{tabular}{|c|c|c|c|c|}
\hline
  $\eps$ &  $N_+^{1k}$ & $N_-^{1k}$&$N_+^{10k}$&$N_-^{10k}$\\\hline\hline
0.04& 608 & 6 & 5939&43\\\hline
0.04& 779 & 968 &7737&9722\\\hline
0.04& 289 & 513 &3131&5115\\\hline
0.04& 749 & 597 &7531&6093\\\hline
0.04& 42 & 378 &315&3745\\\hline
\end{tabular}
\end{minipage}
\end{table}
\end{center}
In table II,  the galactic coordinates (in radians, latitude from the North Galactic Pole) of the annuli with the most significant negative slopes are given for inner-radius 0.01 and widths 0.02 and 0.03. (Sometimes the same point appears in both tables, as with $(\theta,\phi)=(2.219,0.012)$). The slopes are strikingly large apparently pointing to some novel phenomenon.  

\vspace{-5mm}
\begin{center}
\begin{table}[h]
\caption{galactic coordinates of 10 annuli (Planck)\\ with most negative slopes ($T_{\rm CMB}$/rad)\\
left: $r_1=0.01,\ \eps= 0.02$, right: $r_1=0.01,\ \eps= 0.03$}
\vspace{3mm}
\begin{minipage}[b]{0.35\linewidth}\centering
\begin{tabular}{|c|c|c|}
\hline
  $T_{\rm CMB}$/rad& $\theta$ &  $\phi$\\\hline\hline
-0.01403& 2.219& 0.012\\\hline
-0.01266&0.204& 2.405\\\hline
-0.01215&0.703& 2.777\\\hline
-0.01166&2.988& 4.908\\\hline
-0.01164&2.545&0.051\\\hline
-0.01154&2.949&0.052\\\hline
-0.01151&2.678&5.388\\\hline
-0.01129&0.716& 3.698\\\hline
-0.01113&0.689&0.844\\\hline
-0.01103&0.799&0.785\\\hline
\end{tabular}
\end{minipage}
\hspace{8mm}
\begin{minipage}[b]{0.35\linewidth}\centering
\begin{tabular}{|c|c|c|}
\hline
  $T_{\rm CMB}$/rad& $\theta$ &  $\phi$\\\hline\hline
-0.01033&0.204& 2.405\\\hline
-0.00891& 2.962& 0.056\\\hline
-0.00804&2.219& 0.012\\\hline
-0.00735&0.140& 4.783\\\hline
-0.00721&2.383& 0.744\\\hline
-0.00701&2.795&2.705\\\hline
-0.00677&0.799&0.785\\\hline
-0.00662& 2.949& 0.052\\\hline
-0.00660&2.756&5.209\\\hline
-0.00658&2.545&0.051\\\hline
\end{tabular}
\end{minipage}
\end{table}
\end{center}
The difference between the temperatures of the outer and inner boundaries for the most significant annulus $(0.01,0.02)$ is $-2.8\cdot 10^{-4}$ K and for $(0.01,0.03)$ it is $-3.1\cdot 10^{-4}$ K i.e. more than an order of magnitude more than the average CMB fluctuation.

To confirm the result we also used 1000 maps from FFP8.1 MC simulations (with $l\leqslant 1500$) produced and made available by the PLANCK team (Planck Collaboration 2016). These maps put many factors into consideration, such as beam, noise, scanning strategy and gravitational lensing. The simulations have given the results very similar to those given in Table I and are given below in Table III. In particular for $(0.01,0.02)$ and $(0.01,0.03)$ there were no (or 1) simulated maps that outperformed the real map. In addition, we can see that for $(0.00,0.04)$ the number is also significantly low. Therefore, we can draw the conclusion that our previous result was not influenced by gravitational lensing or noise.

\begin{center}
\begin{table}[h]
\caption{\# of artificial FFP8.1 MC maps outperforming the real Planck 70 GHz map}
\begin{tabular}{|c||c|c|c||c|c|c||c|c|c||}
\hline
  $r_1$& $\eps$ &  $N_+^{1k}$ & $N_-^{1k}$&$\eps$ &  $N_+^{1k}$ & $N_-^{1k}$& $\eps$ &  $N_+^{1k}$ & $N_-^{1k}$\\\hline\hline
0.00&0.02& 733 & 187 &0.03& 665 & 55  & 0.04 &581&4\\\hline
0.01&0.02& 705 & 1 &0.03& 878 & 0& 0.04 &780&974\\\hline
0.02&0.02& 387 & 117 &0.03& 772 & 594 & 0.04&315&540\\\hline
0.03&0.02& 260 & 985 &0.03& 163 & 672 & 0.04&736&635\\\hline
0.04&0.02& 983 & 911 &0.03& 143 & 408 & 0.04&31&373 \\\hline
\end{tabular}
\end{table}
\end{center}

We also compared our result with the WMAP 'deconvolved' 1000 maps with the theoretical Planck power spectrum. The results are given below in Table IV. Again for $(0.01,0.02)$ and $(0.01,0.03)$ there were no simulated maps that outperformed the real map. In table V we give the directions of the most intense annuli from the WMAP data. Note that most intense 5 annuli listed in Table II (from  the Planck data) are also in Table V (from the WMAP data).

\begin{center}
\begin{table}[h]
\caption{\# of artificial maps outperforming the real WMAP 70 GHz map}
\begin{tabular}{|c||c|c|c||c|c|c||c|c|c||}
\hline
  $r_1$& $\eps$ &  $N_+^{1k}$ & $N_-^{1k}$&$\eps$ &  $N_+^{1k}$ & $N_-^{1k}$& $\eps$ &  $N_+^{1k}$ & $N_-^{1k}$\\\hline\hline
0.00&0.02& 636 & 187& 0.03& 690 & 47 &0.04&485&4\\\hline
0.01&0.02& 776 & 0 &0.03& 942 & 0 &0.04&731&973\\\hline
0.02&0.02& 535 & 102 &0.03& 726 & 461 &0.04&442&552 \\\hline
0.03&0.02& 186 & 941 &0.03& 341 & 685 &0.04&642&733 \\\hline
0.04&0.02& 995 & 728 &0.03& 240 & 500&0.04&14&326  \\\hline
\end{tabular}
\end{table}
\end{center}

\vspace{-5mm}
\begin{center}
\begin{table}[h]
\caption{galactic coordinates of annuli (WMAP)\\ with most negative slopes ($T_{\rm CMB}$/rad), $r_1=0.01$}
\vspace{3mm}
\begin{minipage}[b]{0.35\linewidth}\centering
\begin{tabular}{|c|c|c|}
\hline
$ \eps$& $\theta$ &  $\phi$\\\hline\hline
0.02& 0.204& 2.405\\\hline
0.02&0.582& 5.323\\\hline
0.02&2.219& 0.012\\\hline
0.02&0.703& 2.778\\\hline
0.02&2.118&1.853\\\hline
0.02&2.988&4.909\\\hline
\end{tabular}
\end{minipage}
\hspace{-6mm}
\begin{minipage}[b]{0.35\linewidth}\centering
\begin{tabular}{|c|c|c|}
\hline
$ \eps$& $\theta$ &  $\phi$\\\hline\hline
0.02&0.595&3.159\\\hline
0.02&2.962& 0.056\\\hline
0.03&0.204&2.405\\\hline
0.03&2.962&0.056\\\hline
0.03&2.118&1.853\\\hline
0.03&2.219&0.012\\\hline
\end{tabular}
\end{minipage}
\hspace{8mm}
\end{table}
\end{center}

\noindent{\bf Outlook}
\vspace{2mm}

\noindent
It seems to us that our analysis of slopes of the temperature in the CMB maps gives us a significant initial indication of the nature of the anomalous regions and provides an important new input into cosmology, irrespective of the validity of CCC. It is hard to see, however, that they find a natural explanation in the currently conventional inflationary picture.  

\vspace{2mm}
\noindent
 {\bf Acknowledgments:}  We thank Pawe{\l} Bielewicz, Arthur Kosowsky, Don Page and Robert Wald for discussions. R.P. thanks J.P. Moussouris for financial assistance through a personal endowment. K.A.M. was partially supported by the Polish National Science Centre grant DEC-2017/25/B/ST2/00165, and P. N. was partially supported by the Polish National Science Centre grant DEC-2019/34/H/ST1/00636. We acknowledge the help from the {\'S}wierk Computing Centre in the National Centre for Nuclear Research (NCBJ, Otwock, Poland).
\vspace{0.8cm}

\centerline{REFERENCES}

\noindent
An, D., Meissner, K.A. and Nurowski, P., 2018,  {\it Ring Type Structures in the Planck map of the CMB}, Mon.Not.Roy.Astron.Soc. {\bf 473},  no.3, 3251.\\[1mm] DeAbreu, A., Contreras, C., Scott, D., 2015, {\it  Searching
       for concentric low variance circles in the cosmic
       microwave background}, JCAP {\bf 12}, 031.\\[1mm]
Gasperini, M. and Veneziano, G., 1993, {\it The pre-big bang
  scenario in string cosmology}, Astropart. Phys. 1, 317\\[1mm]
Gurzadyan, V.G. and Penrose R., 2013, {\it On CCC-predicted concentric low-variance circles in the CMB sky}, Eur. Phys. J. Plus {\bf 128},  22.\\[1mm]
Gurzadyan,V.G. and Penrose, R., 2016, {\it CCC and the Fermi
  paradox}, Eur.Phys.J.Plus {\bf 131}, 11.\\[1mm]
A. Hajian, 2011, {\it Are there echoes from the pre-Big-Bang Universe? A search for low-variance circles in the cosmic microwave background sky}, ApJ {\bf 740}, 52.\\[1mm]
Hawking, S. W., 1974, {\it Black hole explosions?} Nature, {\bf 248}, 30.\\[1mm]
Hawking, S. W., 1975, {\it Particle creation by black holes}, Comm. Math. Phys. {\bf 43}, 199.\\[1mm]
Meissner, K.A., 2012, {\it A Tail Sensitive Test for Cumulative Distribution Functions}, arXiv:1206.4000 [math.ST].\\[1mm]
Meissner, K. A., Nurowski, P. and Ruszczycki, B., 2013, {\it Structures in the microwave background radiation}, Proc. R. Soc. {\bf A469}, 2155.\\[1mm]
Nelson W. and Wilson-Ewing E., (2011) {\it Pre-big-bang cosmology and circles in the cosmic microwave background}, Phys.Rev. {\bf D84} 043508.\\[1mm]
Page, D. N., 1976, {\it Particle emission rates from a black hole: Massless particles from an uncharged, nonrotating hole}, Phys. Rev. {\bf D13} 198.\\[1mm]
Penrose, R., 1964, {\it Conformal approach to infinity, in Relativity, Groups and Topology: The 1963 Les Houches Lectures}, eds. B.S. DeWitt and C.M. DeWitt (Gordon and Breach, New York) \\[1mm]
Penrose, R., 1965, {\it Zero rest-mass fields including gravitation: asymptotic behaviour}, Proc. Roy. Soc. London Ser.A {\bf 284} (1965) 159\\[1mm]
Penrose, R., 2006, {\it Before the big bang: an outrageous new perspective and its implications for particle physics}. In EPAC 2006 Proc., Edinburgh, Scotland, pp.2759-62 ed. C.R.Prior.\\[1mm]
Penrose, R., 2010, {\it Cycles of Time: An Extraordinary New View of the Universe}, Bodley Head,  London.\\[1mm]
Penrose, R., 2018, {\it The Big Bang and its Dark Matter
      Content: Whence, Whither, and Wherefore},
Found. Phys. {\bf 6}, 1.\\[1mm]
Planck Collaboration,  2016, Astron. Astrophys. {\bf 594}, A12, arXiv:1509.06348.\\[1mm]
Tod, K. P., 2012, {\it Penrose’s circles in the CMB and a test of inflation}, Gen. Relativ. Gravit. {\bf 44} 2933.\\[1mm]
Wehus, I. K. and Eriksen H. K., 2011, {\it A search for concentric circles in the 7 year Wilkinson Microwave Anisotropy Probe temperature sky maps}, ApJ {\bf 733}, L29 

\end{document}